\documentclass{ws-ijmpa}

\usepackage[dvips]{color}
\newcommand{\beq}{\begin{equation}}
\newcommand{\eeq}{\end{equation}}
\newcommand{\bq}{\begin{equation}}
\newcommand{\eq}{\end{equation}}
\newcommand{\ba}{\begin{array}}
\newcommand{\ea}{\end{array}}
\newcommand{\beqa}{\begin{eqnarray}}
\newcommand{\eeqa}{\end{eqnarray}}

\def\R{{\cal R}}

\def\End{\end{document}}

\def\to{\rightarrow}

\def\dis{\displaystyle}
\def\f{\frac}
\def\ov{\overline}

\def\[{\left[}
\def\]{\right]}
\def\({\left(}
\def\){\right)}

\def\l{{\ell}}

\def\U1EM{U(1)_{\rm em}}

\def\R{\mathcal R}

\def\leqq{\leqslant}
\def\geqq{\geqslant}

\def\d{\delta}

\def\[{\left[}
\def\]{\right]}
\def\dis{\displaystyle}

\def\cut{\Lambda}

\def\Eu{E^{\star}}

\def\nL{\nu_L^{~}}

\newcommand{\bla}{\color{black}}
\newcommand{\red}{\color{red}}
\newcommand{\blu}{\color{blue}}

\begin{document}

\markboth{H.-J. He and D. A. Dicus}
{Scale of Mass Generation for Majorana Neutrinos}

%
\catchline{}{}{}{}{}
%

\title{SCALE OF MASS GENERATION FOR MAJORANA NEUTRINOS
$\!$\footnote{Talk presented by H.J.H. at the Conference of
the Division of Particles and Fields, American Physical 
Society (DPF-2004), held at University of California, 
Riverside, CA, USA, August 26-31, 2004. 
}
}

\author{\normalsize {\sc Hong-Jian He}    \,~~and~~
                    {\sc Duane A. Dicus}}

\address{\small Department of Physics, University of Texas at Austin, 
                TX 78712, USA}

%

\maketitle

\pub{Received: 1 November, 2004}{} 

\begin{abstract}
Scales of mass generation 
for Majorana neutrinos (as well as quarks and leptons)
can be probed from high energy  
\,$2\to n$\, inelastic scattering involving
a multiple longitudinal gauge boson final state.  
We demonstrate that the unitarity of \,$2\to n$\, scattering
puts the strongest new upper limit on the scale
of fermion mass generation, independent of the electroweak
symmetry breaking scale \,$v=(\sqrt{2}G_F)^{-\f{1}{2}}$.\,
Strikingly, for Majorana neutrinos (quarks and leptons), 
we find that the strongest
$2\to n$ limits fall in a narrow range, 
$136-170$\,TeV ($3-107$\,TeV)
with $n=20-24$ ($n=2-12$), depending on the observed
fermion masses. Physical implications are discussed.

\end{abstract}

\section{Scale of Fermion Mass Generation: the Puzzle}

The standard model (SM) hypothesizes
a fundamental Higgs boson to generate masses for
all elementary particles: weak gauge bosons, quarks and leptons.
The masses $M_{W,Z}$ of gauge bosons $(W^\pm,Z^0)$ involve
the electroweak gauge couplings $(g,\,g')$ and the vacuum expectation
value $v=(\sqrt{2}G_F)^{-\f{1}{2}}$, while the quark/lepton masses
arise from the product of Yukawa couplings $y_f^{~}$ and $v$\,.\, 
Unlike gauge interaction, the flavor-dependent 
Yukawa couplings $y_f^{~}$ 
are completely arbitrary,
exhibiting a large hierarchy 
\,$y_e^{~}/y_t^{~}=m_e^{~}/m_t^{~} \simeq 3\times10^{-6}$\,
between the electron and top quark.
No compelling principle requires the fermion mass generation to
share the same mechanism as the $W/Z$ gauge bosons.
Moreover, the tiny neutrino masses $\,m_\nu^{~}\gtrsim 0.05\,$eV
could not be generated by such a Higgs boson without
losing renormalizability\cite{wein5} or extending the SM
particle spectrum\cite{nu-seesaw,nu-rad}.
So far, neither is the Higgs boson found nor is any
Yukawa coupling experimentally measured --- the origin and scale of
mass generation remain, perhaps, the greatest mystery.

What is wrong with just putting all the {\it bare masses} into 
the SM Lagrangian by hand? These bare mass-terms can be made gauge
invariant in the nonlinear realization\cite{CCWZ},
but are manifestly nonrenormalizable. 
This leads to unitarity violation in
high energy scattering at a scale $\Eu$. 
Generically, we define the scale 
$\cut_x$ for generating a mass $m_x^{~}$ to be the {\it minimal energy
above which the bare mass term for $m_x^{~}$ has to be replaced 
by a renormalizable interaction} (adding at least one new physical
state to the particle spectrum). Hence the unitarity violation
scale $\Eu$ puts a universal upper bound on $\cut_x$, i.e.,
\,$\cut_x \leqq \Eu$\,.\,

A mass term for the bare masses $M_{W,Z}$ will cause 
the high energy $2\to 2$ longitudinal gauge boson scattering 
to violate unitarity at a scale,\cite{DM,LQT,Vel,CG}
\beq
\,E^\star_W ~\simeq~ \sqrt{8\pi}v ~\simeq~ 1.2\,{\rm TeV}\,.
\eeq
This puts an upper limit on the scale of the electroweak
symmetry breaking (EWSB) and justifies TeV energy scale for 
the construction of the CERN LHC. 
Similarly, with the bare mass terms for 
Dirac fermions (quarks and leptons), 
an upper bound on the scale of fermion mass generation 
can be derived from the \,$2\to 2$\, inelastic scattering  
$f_\pm\bar{f}_\pm \to V_L^{a_1}V_L^{a_2}$ 
($V^a=W^\pm,Z^0$),\cite{AC}
\beq
\label{eq:UBf22}
E_f^\star ~\simeq~ \dis  \f{8\pi v^2}{~\sqrt{N_c}\,m_f^{~}~} \,,
\eeq
where $N_c = 3\,(1)$ for quarks (leptons).
This shows that the upper limit on
the scale of fermion mass generation is proportional to 
$1/m_f^{~}$ and is thus independent of the classic 
bound $E^\star_W$ on the EWSB scale.
Also, for all the SM fermions (except the
top quark), the limit (\ref{eq:UBf22})
is substantially higher than $E^\star_W$.
For the scale of mass generation for Majorana neutrinos, 
Refs.\,[\refcite{scott-PRL,scott}] derived
an analogous $2\to 2$ bound, 
\beq
\label{eq:UBnu22}
E_\nu^\star ~\simeq~ \dis  
\f{~\,4\pi v^2\,}{\,~m_\nu^{~}~} \,,
\eeq
which is about \,$10^{16}$\,GeV\, (the GUT scale) for 
typical input \,$m_\nu^{~}\sim 0.05$\,eV [\refcite{superK}].

However, by considering the $2\to n$ inelastic scattering,
$f\bar{f},\nu\nu \to nV_L^a$ ($n>2$), 
it was noted\cite{scott} that the $n$-body phase space
integration contributes a large energy power factor $s^{n-2}$ 
to further enhance the cross section 
(in addition to the energy dependence $s^1$ 
from the squared amplitude), so the unitarity limit for 
a fermion $f\,(\nu)$ would behave like
\beq
\label{eq:UBfscot}
E_{f(\nu)}^\star ~\sim~ \dis 
v\(\f{v}{\,m_{f(\nu)\,}^{~}}\)^{\f{1}{n-1}}
\,\longrightarrow~\, v\,,~~~~~({\rm for}\,~n\to\,{\rm large})\,,
\eeq
which could be pushed arbitrarily close to the weak scale $v$ and
thus become independent of the fermion mass 
$m_f^{~}$ for large enough $n$.
This raises a deep puzzle: {\it is there an independent new scale
for fermion mass generation revealed from the fermion-(anti)fermion
scattering into weak bosons?}  
We find the behavior (\ref{eq:UBfscot}) very counter-intuitive
since the kinematic condition forces
any \,$2\to n$\, unitarity limit $\Eu$
to grow at least linearly with $n$ [\refcite{DH}],
\beq
\label{eq:KC}
\dis
\sqrt{s} ~\,>~\, n M_{W(Z)} \,\simeq\, \dis\f{n}{3}v ~,
~~~\longrightarrow~~~
\f{\,E^\star}{v} ~\,>~\, \f{n}{3} ~.
\eeq

\section{The Resolution and New Upper Limits}

\begin{figure}[t]
\label{fig:Fig1}
\begin{center}
\centerline{\psfig{file=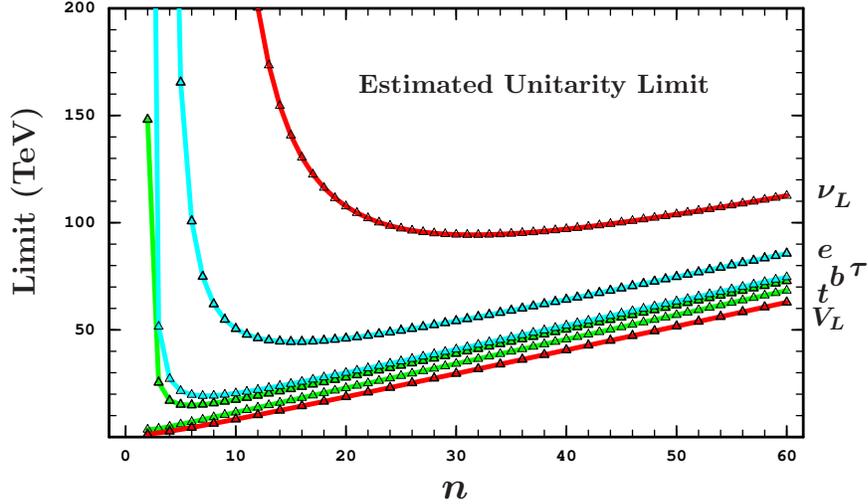,width=10cm}}
\setlength{\unitlength}{1mm}
\begin{picture}(0,0)
\put(51,40){\large$\mbox{\boldmath$\nL$}$}
\put(51,32){\large$\mbox{\boldmath$e$}$}
\put(55.3,29.6){\large$\mbox{\boldmath$\tau$}$}
\put(52.6,28.3){\large$\mbox{\boldmath$b$}$}
\put(51,26.1){\large$\mbox{\boldmath$t$}$}
\put(50.2,23){$\mbox{\boldmath$V_L$}$}
\put(-10,54){\bf{\normalsize Estimated Unitarity Limit}}
\put(1,0){\Large $\mbox{\boldmath$n$}$}
\put(-52,27){\rotatebox{90}{\makebox(0,0)[lb]{\bf{\large Limit (TeV)}}}}
\end{picture}
\caption{Realistic estimate of the unitarity bound $E^\star$
for the scattering processes 
\,$V_L^{a_1}V_L^{a_2},\,t\bar{t},\,b\bar{b},\,$ $
   \tau^-\tau^+,\,e^-e^+,\,\nL\nL
   \to nV_L^a$\,
(curves from bottom to top),
as a function of \,$n\,(\geqq 2)$\,.}
\end{center}
\end{figure}

\subsection{The Resolution}

We observe that the resolution to this puzzle has to rely on the 
additional $n$-dependent {\it dimensionless} factors in the 
exact n-body phase space integration which can sufficiently
suppress the $E$-power enhancement $s^{n-2}$ mentioned above
Eq.\,(\ref{eq:UBfscot}).
Computing the exact $n$-body phase space we can 
formally derive the $2\to n$ unitarity limit 
as follows,\cite{DH} 
\beq
\label{eq:UBus}
\dis
E^\star ~=~ v\[C_0\, 2^{4n-2}\pi^{2(n-1)}
            \varrho\, (n-1)!\,(n-2)!
\(2N_c\)^{\d-2}
\(\f{v}{\,m_{f,\nu}^{~}\,}\)^{2(2-\d)}\]^{\f{1}{2(n-2+\d)}}
\eeq
where $\,\d = 1\,(2)$ for the scattering 
\,$f\bar{f},\nL\nL (V_L^{a_1}V_L^{a_2})\to nV_L^a$\,,\,
$\varrho$ is the symmetry factor from the 
identical particles in the final state,
and the constant $\,C_0$\, originates from the dimensionless
coefficient in the squared amplitude (whose possible angular
dependence is included in the exact $n$-body phase 
space integration). Since all large $n$ factors are explicitly
counted in (\ref{eq:UBus}),
we expect that, independent of the detail, 
$~C_0^{\f{1}{\,2(n-2+\d)\,}} ~\sim~ 1~$ reasonably holds and
becomes increasingly more accurate as $n$ gets larger.
So, setting $C_0\simeq 1$ and $\varrho\simeq 1$ for simplicity we
deduce an realistic estimate of the unitarity limit \,$\Eu$\,
shown in Fig.\,1 (where $\,m_{\nu}^{~}=0.05$\,eV\, is chosen).   
Using the Stirling formula 
$\,n!\simeq n^ne^{-n}\sqrt{2\pi n}\,$, we can then derive the
correct asymptotic behavior from (\ref{eq:UBus}),
\beq
\label{eq:asymp}
\dis
\Eu \,\to\, v\f{4\pi n}{e} ~~>~~ v\f{n}{3} \,,
~~~~~~~~({\rm for}~n\gg 1)\,, 
\eeq
in full agreement with the kinematic condition (\ref{eq:KC}).

We see that for \,$V_L^{a_1}V_L^{a_2}$\, 
and \,$t\bar{t}$\, initial states
the strongest bound (minimum of the curve) still occurs at $n= n_s=2$;
while for {\it all light fermions} including Majorana neutrinos
the best limit for the scale of mass generation lies at a {\it new minimum}
with  $\,n= n_s > 2 \,$ and $\,\Eu\,$ no higher than about $100\,$TeV,
which is substantially tighter than 
the corresponding classic $2\to 2$ bound.
As we observed\cite{DH}, it is the {\it competition} between the
large asymptotic linear growth (\ref{eq:asymp}) and the strong
power suppression (\ref{eq:UBfscot}) that generates
a genuine {\it new minimum scale $E^\star_{\min}$} 
for all light fermions
at $\,n= n_s>2\,$, {\it independent of the EWSB scale.}

\subsection{Scale of Mass Generation for Majorana Neutrinos}

\begin{figure}[t]
\label{fig:Fig2}
\begin{center}
\centerline{\psfig{file=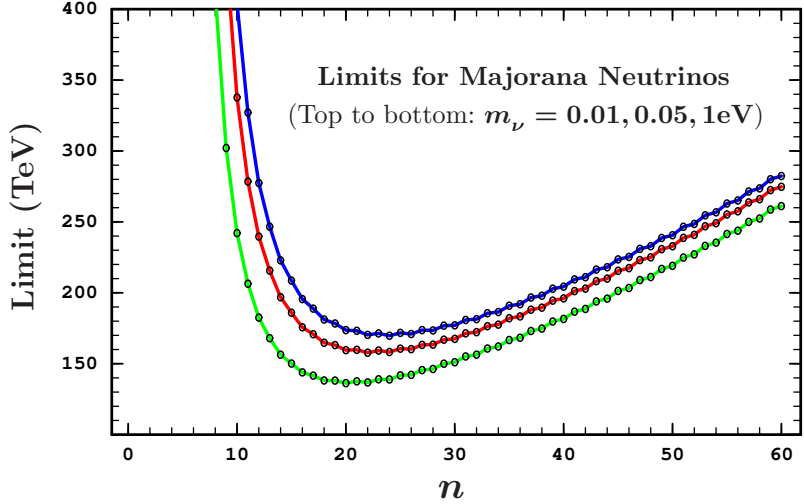,width=10cm}}
\setlength{\unitlength}{1mm}
\begin{picture}(0,0)
\put(-15,55){\bf{\normalsize Limits for Majorana Neutrinos}}
\put(-19,50){(Top to bottom: 
$\mbox{\boldmath$m_\nu^{~}=0.01,0.05,1$}${\bf eV})}
\put(1,0){\Large $\mbox{\boldmath$n$}$}
\put(-51.7,25){\rotatebox{90}{\makebox(0,0)[lb]{\bf{\large Limit (TeV)}}}}
\end{picture}
\caption{Precise unitarity limit $E^\star_\nu$ of Majorana neutrinos
from the scattering processes 
\,$\nL\nL \to nV_L^a$\, (curves from bottom to top),
as a function of \,$n\,(\geqq 2)$\,.}
\end{center}
\end{figure}

For Majorana neutrinos, 
the dimension-3 bare mass term can be written as
$\,-\f{1}{2}m_{\nu}^{ij}\nu_{Li}^T\widehat{C}\nu_{Lj}+{\rm H.c.},\,$
which takes the gauge-invariant nonlinear form 
\beq
\label{eq:nu-mass-pi}
{\cal L}_\nu ~=~ \dis
-\f{1}{2}m_\nu^{ij}{L^{\alpha}_i}^T \widehat{C}L^{\beta}_j
 \ov{\Phi}^{\alpha'}\ov{\Phi}^{\beta'}
\epsilon^{\alpha\alpha'} \epsilon^{\beta\beta'} + {\rm H.c.}
\,,
\eeq
where $\,\widehat{C}=i\gamma^2\gamma^0\,$,\,
$\,\ov{\Phi}=U(0,1)^T$\,,\, $U=\exp[i\pi^a\tau^a/v]$\,
and $\,L_j=(\nu_{Lj}^{~}/\sqrt{2},\,\ell_{Lj})^T\,$.\,
We then quantitatively compute the high energy scattering
\,$\nL\nL\to nV_L^a\,(n\pi^a)$ by
using the equivalence theorem\cite{LQT,CG,ET}. 
The leading amplitude \,$\nL\nL\to n\pi^a$\, 
is given by the $\nL$-$\nL$-$\pi^n$ contact interactions 
in (\ref{eq:nu-mass-pi}) and 
is of $\,O(m_{\nu}^{~}E/v^n)\,$
by power counting\cite{PC}.
Defining a real Majorana neutrino field
$\,\chi =(\nL +\nu_L^c)/\sqrt{2}\,$,  
we compute the scattering amplitude of
$\,\f{1}{\sqrt{2}}\[
|\chi_{j+}\ov{\chi}_{j+}\rangle \mp
|\chi_{j-}\ov{\chi}_{j-}\rangle \] \to
\(\pi^+\)^{\ell}\(\pi^-\)^{\ell}\(\pi^0\)^{n-2\ell}
\,$
for three cases: 
(a). $n({\rm even})=2\l$\,;\,   
(b). $n({\rm even}) > 2\l$\,;\,
(c). $n({\rm odd})\geqq 3$\,.\,
Thus, we derive the unitarity limit,\cite{DH}
\beqa
\label{eq:UBnu-final}
E^\star_\nu ~~&=&~~ v\[\(\f{v}{m_{\nu j}^{~}}\)^2
                       \f{4\pi}{\,\R_{\nu k}^{\max}\,}
             \,\]^{\f{1}{2(n-1)}}  ,
\eeqa
where $~\R_{\nu k}^{\max}$~ is given by   
\beq
\label{eq:Rmax-nu}
\ba{lcll}
\R_{\nu 1}^{\max} &=&
\dis\f{\(\f{n}{2}!\)^2 (K_{1,n})^2
}{~2^{3n-4}\pi^{2n-3}\,(n-1)!\,(n-2)!~} \,,
&~~~~~(n={\rm even},~\ell=\f{n}{2})\,,
\\[4mm]
\R_{\nu 2}^{\max} &=&
\dis\f{~n\(K_{1,n}+K_{3,n}\)^2~}
{~2^{4(n-1)}\pi^{2n-3}\,(n-2)!~} \,,
&~~~~~(n={\rm even},~\ell < \f{n}{2})\,,
\\[4mm]
\R_{\nu 3}^{\max} &=&
\dis\f{~n\,(K_{2,n})^2~}{~2^{4(n-1)}\pi^{2n-3}\,(n-2)!~} \,,
&~~~~~(n={\rm odd})\,,
\ea            
\eeq
and \,$K_{j,n}$\, are functions of $n$ [\refcite{DH}].
For \,$n=$~even\,,\, the best limits are from
$\,\(\R_{\nu 1}^{\max},\,\R_{\nu 2}^{\max}\)_{\max} $,\, but 
numerically  the bounds from
\,$\R_{\nu 1}^{\max}$\, and \,$\R_{\nu 2}^{\max}$\,
only differ by less than 2\%.\,  As shown in Fig.\,2, we find
the optimal limits, 
\beq
\label{eq:nuB-min}
E^\star_{\min} = 136,\,158,\,170\,{\rm TeV},~~~~~~
{\rm located~at}~~n=n_s = 20,\,22,\,24,
\eeq
for \,$m_{\nu j}^{~} = 1,\, 0.05,\, 0.01$\,eV,\, respectively.
This agrees with our  
estimate in Fig.\,1 to less than
about a factor of \,$1.7$\,.

\subsection{Scale of Mass Generation for Quarks and Leptons}

Similarly, we derive the quantitative \,$2\to n$\, bounds for 
Dirac fermions --- the quarks and leptons.
Consider a pair of SM fermions $(f,f')$ which forms a left-handed
$SU(2)_L$ doublet $F_L=(f_L,f'_L)^T$ and two right-handed weak
singlets $f_R$ and $f_R'$. We then formulate their bare Dirac mass-terms
$\,-m_f^{~}\ov{f}f - m_{f'}^{~}\ov{f'}f'\,$ 
into the gauge-invariant nonlinear form,
\beq
\label{eq:f-mass}
{\cal L}_f = \dis
-m_f\ov{F_L}U\(\ba{c} 1 \\ 0 \ea\)f_R
-m_{f'}\ov{F_L}U\(\ba{c} 0 \\ 1 \ea\)f'_R + {\rm H.c.}
\eeq 
Thus we compute the scattering amplitude for 
$\,|{\rm in}\rangle \to (\pi^+)^k(\pi^-)^{\ell}(\pi^0)^{n-k-\ell}\,$,\,
where the initial state $|{\rm in}\rangle$ consists of two fermions
in the color-singlet channel\cite{DH}.\,
From a systematical analysis we derive the unitarity limit
for quarks and leptons,
\beqa
\label{eq:UBf-final}
E^\star_f ~~&=&~~ v\[\(\f{v}{m_{\widehat{f}}^{~}}\)^2
               \f{4\pi}{\,N_c\,\widehat{\R}_j^{\max}\,}
               \]^{\f{1}{2(n-1)}}  ,
\eeqa
where $\,\widehat{\R}^{\max}_j =
         \(2\R_1^{\max},\,2\R_2^{\max},\,\R_3^{\max}\)\,$ 
is a function of $\,n\,$  [\refcite{DH}]. 
We depict the numerical bounds $\,E_f^{\star}\,$ in Fig.\,3,
which shows that the best limits for all light fermions
occur at a {\it new minimum} $\,n= n_s>2\,$
with $\,E^\star_b=24.5\,$TeV for $b$ quark and 
     $\,E^\star_e=107\,$TeV for electron. 
These two limits only differ by about a factor $4$\,.

\begin{figure}[t]
\label{fig:Fig3}
\begin{center}
\centerline{\psfig{file=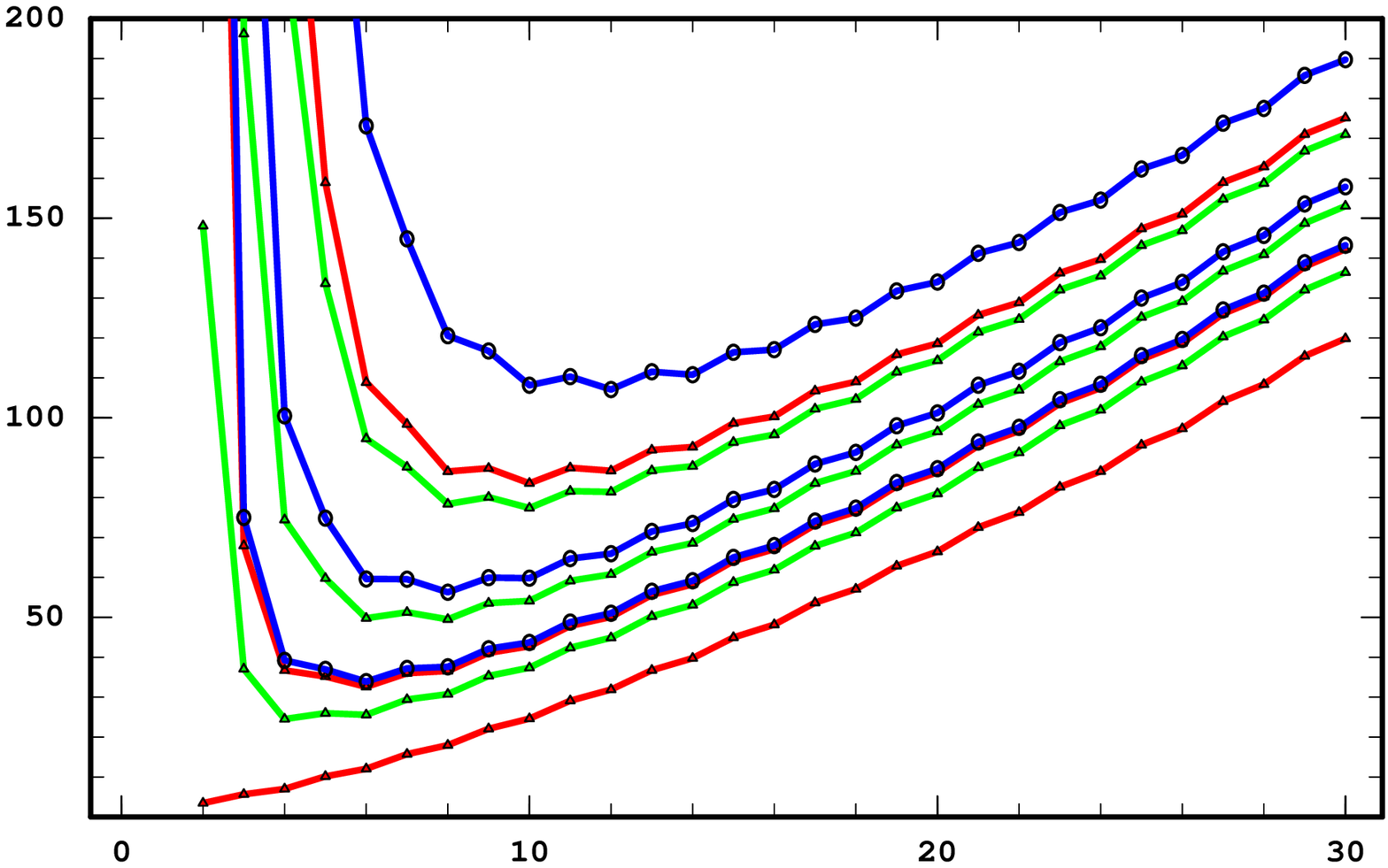,width=10.5cm}}
\setlength{\unitlength}{1mm}
\begin{picture}(0,0)
\put(53,65){\large$\mathbf{e}$}
\put(53,60.5){\large$\mathbf{u}$}
\put(55.8,59){\large$\mathbf{d}$}
\put(53,56.5){\large$\mbox{\boldmath$\mu$}$}
\put(56.2,53.8){\large$\mathbf{s}$}
\put(53,51.6){\large$\mbox{\boldmath$\tau$}$}
\put(56.2,50){\large$\mathbf{c}$}
\put(53,48){\large$\mathbf{b}$}
\put(53,44){\large$\mathbf{t}$}
\put(-18.5,60.4){\bf{Limits for Quarks \& Leptons}}
\put(3,4){\Large $\mbox{\boldmath$n$}$}
\put(-50,27){\rotatebox{90}{\makebox(0,0)[lb]{\bf{\large Limit (TeV)}}}}
\end{picture}
\caption{Precise unitarity limits $E^\star_f$ of
quarks and leptons
for the scattering  
$\,t\bar{t},\,b\bar{b},\,c\bar{c},\,\tau^-\tau^+,$ $\,
   s\bar{s},\,\mu^-\mu^+,\,d\bar{d},\,u\bar{u},\,e^-e^+
\to nV_L^a$\,
(curves from bottom to top),
as a function of \,$n\,(\geqq 2)$\,.}
\end{center}
\vspace*{-3mm}
\end{figure}

\section{Discussions and Conclusions}

We have systematically analyzed the unitarity limits
from the $2\to n$ inelastic scattering ($n\geqq 2$), 
which provide a universal upper bound 
on the scale of mass generations  
for Majorana neutrinos as well as leptons and quarks. 
The numerical results\cite{DH} are summarized by Table\,1, in 
comparison with the classic $2\to 2$ limits\,\cite{AC,scott}.

\begin{table}[h]
\tbl{Summary of our best unitarity limit $E^{\star}_{2\to n}$ (in TeV)
from each scattering $\,\xi^{~}_1\xi^{~}_2\to nV_L^a\,(n\pi^a)$\,,\,
in comparison with
the classic $2\to 2$ limit  $E^{\star}_{2\to 2}$ (in TeV).}
%
{\begin{tabular}{c||c|cccccc|ccc|c}
\hline\hline\hline
&&&&&&&&&&&\\[-2.5mm]
   $\blu\xi_1^{~}\xi_2^{~}$      
&  $\blu V_L^{a_1}V_L^{a_2}$       &    $\blu t\ov{t}$     &   
$\blu b\ov{b}$     & $\blu c\ov{c}$  & $\blu s\ov{s}$     &
$\blu d\bar{d}$    & $\blu u\ov{u}$  & $\blu\tau^-\tau^+$ & 
$\blu\mu^-\mu^+$   & $\blu e^-e^+$   & ~~$\blu\nu\nu$~~    
 \\[1.5mm]
\hline\hline
&&&&&&&&&&&\\[-2.5mm]
$\red n_s$ 
&  $\red 2$  & $\red 2$   & $\red 4$ & $\red 6$ & 
$\red 8$   &   $\red 10$ & $\red 10$ ~& $\red 6$ & $\red 8$ & 
$\red 12$  &   $\red 22$  
\\ [1.5mm]   
\hline
&&&&&&&&&&&\\[-2.5mm] \red
$E^{\star}_{2\to n}$ & $\red 1.2$ 
&  $\red 3.5$ & $\red 25$ & $\red 33$   & $\red 49$ 
&  $\red 77$  & $\red 84$~
&  $\red 34$  & $\red 56$ & $\red 107$  & $\red 158$   
\\[1.5mm]
\hline\hline 
&&&&&&&&&&&\\[-2.5mm]\bla
$E^{\star}_{2\to 2}$
& $1.2$  & $3.5$         & $148$ & $497$ & $4\!\!\times\!\!10^3$ 
& $10^5$ & $2\!\!\times\!\!10^5$ & $605$ & $10^4$ 
& $2\!\!\times\!\! 10^6$ & $10^{13}$  
\\[1.5mm]
\hline\hline\hline
\end{tabular}}
\end{table}

Table\,1 shows that the scattering
$\,f\bar{f}\to nV_L^a\,$ ($n\geqq 2$) does reveal
an {\it independent scale} for the fermion mass generation.
Our new unitarity bounds from the $2\to n$ scattering with
$\,n>2\,$ establish a {\it new scale of mass generation} for all
light fermions including Majorana neutrinos and are substantially
stronger than the classic $2\to 2$ limits.
In particular, for the Majorana neutrinos with typical
mass values $\,m_{\nu}^{~}=1-0.01\,$eV 
(as suggested by the current 
experiments on neutrino oscillations, 
neutrinoless double-beta decay 
and astrophysics constraints), the best upper limits on the
scale of mass generation are in the range $136-170$\,TeV
(with $n= n_s=20-24$). This is only about a factor \,$7$\,
weaker than the lowest bound for 
the light Dirac fermions (the $b$ quark)
despite their huge mass hierarchy
$\,m_{\nu}^{~}/m_b^{~}\approx 2\times(10^{-10}-10^{-12})\,$.
Hence, these limits are very insensitive to the variation of 
fermion masses. 
Such a strong non-decoupling feature for the scale of new physics 
associated with light fermion mass generation is essentially due to
the {\it chiral structure} of the bare fermion mass-terms, i.e., the 
fact that all the left-handed SM fermions are weak-doublets but their
right-handed chiral partners are weak singlets (or possibly absent in
the case of Majorana neutrinos with radiative mass generation),
so the decoupling theorem\cite{DT} no longer applies. 
Finally, we have also estimated 
the \,$2\to n$\, unitarity limit on 
the electroweak symmetry breaking (EWSB) scale  
via the scattering 
$V_L^{a_1} V_L^{a_2} \to n V_L^a$ ($\pi^{a_1} \pi^{a_2} \to n \pi^a$)
with $n\geqq 2$, and find that the best limit   
remains to be \,$E^{\star}_W \simeq 1.2$\,TeV 
with $\,n=n_s=2\,$, in agreement with the customary bound.

\vspace*{3mm}
\noindent
{\bf Acknowledgments} \\[1mm]
We thank S. Willenbrock for valuable discussions.
This work was supported by U.S. Department of Energy under
grant No.~DE-FG03-93ER40757.

\end{document}